\documentclass[reprint,amssymb,amsmath,aip]{revtex4-1}
\usepackage{bm}%
\usepackage{color}
\usepackage{graphicx}

\begin{document}
\title {Ab initio phase diagram of BaTiO$_3$ under epitaxial strain revisited}%
\author{Anna Gr\"unebohm}
\email{anna@thp.uni-duisburg.de}
\affiliation{Faculty of Physics and Center for Nanointegration, CeNIDE, University of Duisburg-Essen, 47048 Duisburg, Germany}
\author{Madhura Marathe}
\author{Claude Ederer}
\affiliation{Materials Theory, ETH Z\"urich, 8093 Z\"urich, Switzerland}

\begin{abstract}
We revisit the phase diagram of BaTiO$_3$ under biaxial strain using a
first principles-based effective Hamiltonian approach. We show that,
in addition to the tetragonal ($c$), quasi-rhombohedral ($r$), and
quasi-orthorhombic ($aa$) ferroelectric phases, that have been
discussed previously, 
there are temperature and strain regions, in particular
under tensile strain, where the system decomposes into multi-domain
structures. In such cases, the strained system, at least on a local
level, recovers the same phase sequence as the unclamped bulk
material. Furthermore, we extend these results from the case of
``uniform'' biaxial strain to the situation where
the two in-plane lattice constants are strained differently and show
that similar considerations apply in this case.
\end{abstract}

\maketitle

The optimization of ferroelectric materials by epitaxial growth and
interface-mediated strain is nowadays a well-established and highly
successful method.~\cite{Schlom} However, the experimental
determination of strain-temperature phase diagrams is quite
challenging, since only specific strain values, corresponding to the
given lattice mismatch with a specific substrate, can be investigated.
Therefore, the theoretical modeling of strain-dependent phase diagrams
is highly relevant.

An important case is the prototypical ferroelectric BaTiO$_3$ (BTO),
which, in its free bulk form, exhibits one paraelectric and three
different ferroelectric structures as function of
temperature,\cite{Toshio} and thus gives rise to a rich strain
dependence. Different levels of sophistication have been used to
model/calculate the strain-dependent phase diagram of BTO, however,
leading in part to conflicting results. First, various calculations
based on Ginzburg-Landau-Devonshire theory have been performed,
yielding qualitatively consistent phase diagrams as long as only
mono-domain phases are taken into account.\cite{pertsev2, choi} Once
multi-domain configurations are considered, different (meta-) stable
domain patterns have been found, depending on the {\it a priori}
assumptions of the models.\cite{pertsev4, pertsev3} More recently,
phase field simulations have been used in order to simulate different
phases and domain structures without {\it a priori}
assumptions.~\cite{Chen:2008}

Both approaches, however, require parameters that have to be obtained
from experimental data, and different choices for these parameters can
lead to significantly different results.~\cite{Dieguez,choi} From this
perspective, first principles-based calculations are very attractive,
since in principle they do not require fitting. 
Indeed, first principles-based strain-temperature
phase diagrams for BTO have been
calculated,~\cite{Dieguez,Lai_et_al:2004,Feram2,Madhura} but only
mono-domain states have been found for simulation cell sizes of about
12-16 unit cells along each cartesian direction.

Apart from determining the stability of different phases, the
mechanical boundary conditions can also modify the domain pattern of
the material. For example, the formation of 90$^\circ$ domain walls
provides an efficient way for tetragonal perovskites to partially
relax the elastic energy under clamping to a periodic
substrate.\cite{Pompe} In PbTiO$_3$, such domain configurations have
been predicted theoretically even without epitaxial
constraints.\cite{Kouser} In contrast, a rather large domain wall
energy of 0.9~meV/\AA$^2$ has been obtained for 90$^{\circ}$ domain
walls in BTO,\cite{Kleemann}
and it has been predicted that such domains are not stable for
unclamped BTO.\cite{Nishimatsu2,Feram1}
Experimentally, both multi-domain and mono-domain phases have been
reported for BTO grown under tensile strain.\cite{Qiao,Misirlioglu,He}

In the present paper, we show the presence of multi-domain states 
within a first principles-based effective Hamiltonian approach for
sufficiently large simulation cells. For uniform biaxial tensile
strain, the paraelectric phase first transforms to a multi-domain
ferroelectric phase with local polarization along $\langle 100
\rangle$ and under further cooling a mono-domain state polarized along
$\langle 110 \rangle$ is obtained. For small strains a multi-domain
state with both in-plane and out-of-plane polarization is observed,
while under compressive strain only mono-domain phases
appear. Qualitatively the same trends are found for non-uniform
biaxial strain, where a large variety of multi-domain states exist.

We use the feram code,~\cite{Feram1} which is
based on the effective Hamiltonian discussed in
Refs.~{\onlinecite{King,Zhong,Feram1}}, and the parametrization given
in Ref.~{\onlinecite{Nishimatsu}}. We perform molecular dynamics
simulations in the canonical ensemble, employing a Nos{\'e}-Poincar\'e
thermostat,~\cite{Nose} periodic boundary conditions, and a simulation
box of 32$\times$32$\times$32 unit cells, i.e.\ corresponding to
periodic images with about 13~nm distance. 
An ideal bulk material is modeled, without
surfaces and depolarizing fields. To simulate the effect of epitaxial
strain, the elements $\eta_1$ and $\eta_2$ of the homogeneous strain
tensor (in standard Voigt notation) are fixed to the external strain
$\eta$, and $\eta_6$ is set to zero. We define $\eta=0$ corresponding
to a lattice constant of 3.996~{\AA}, as obtained within our approach
for the free system directly above T$_C$. In the following, the two
clamped in-plane directions are denoted as $a$ and $b$, and the free
out-of-plane direction as $c$.

\begin{figure}
 \centering \includegraphics[height=0.35\textwidth,clip,trim=.5cm 0cm 3cm .5cm,angle=270]{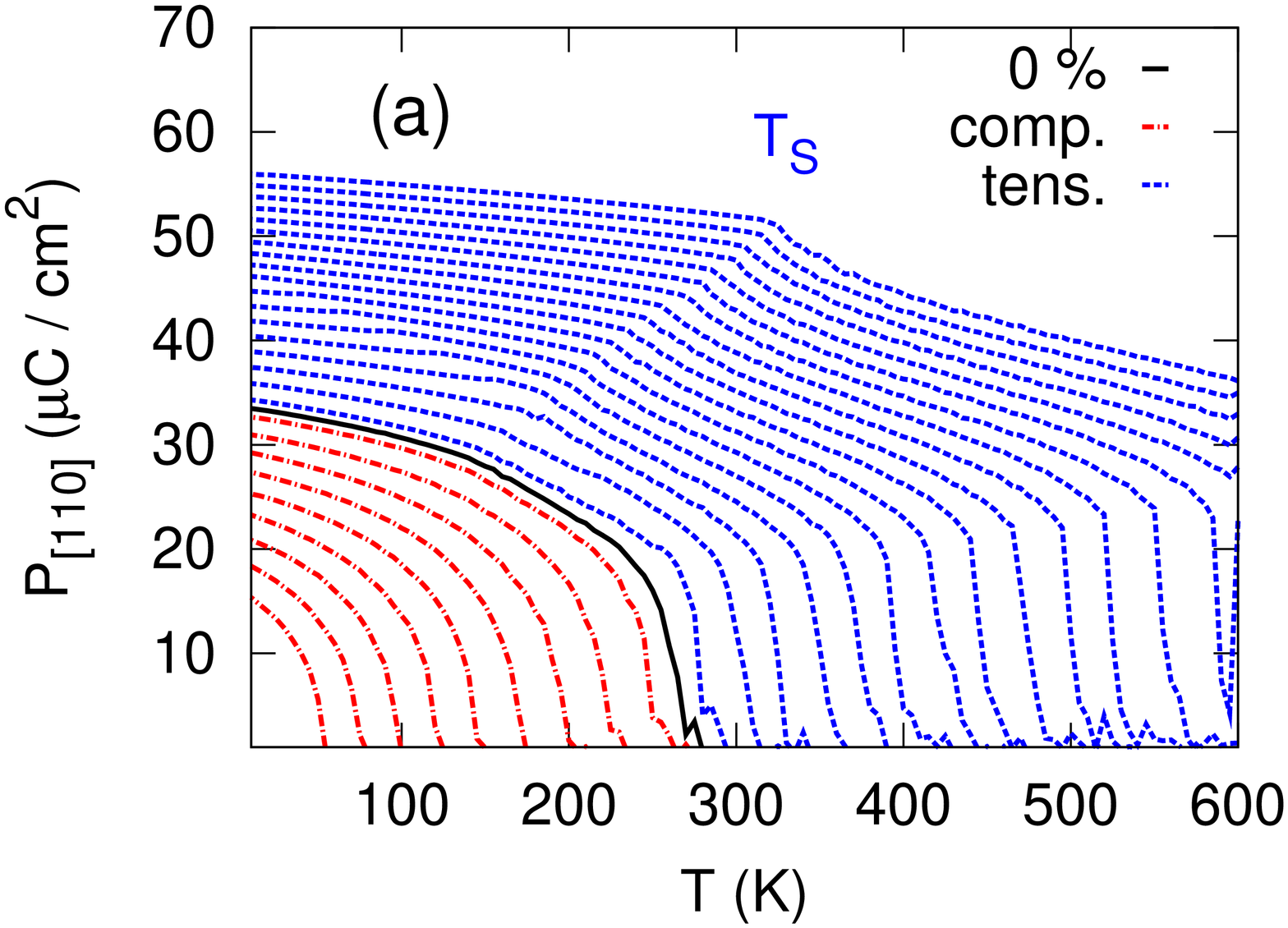}
\includegraphics[height=0.35\textwidth,clip,trim=.5cm 0cm 0cm .5cm, angle=270]{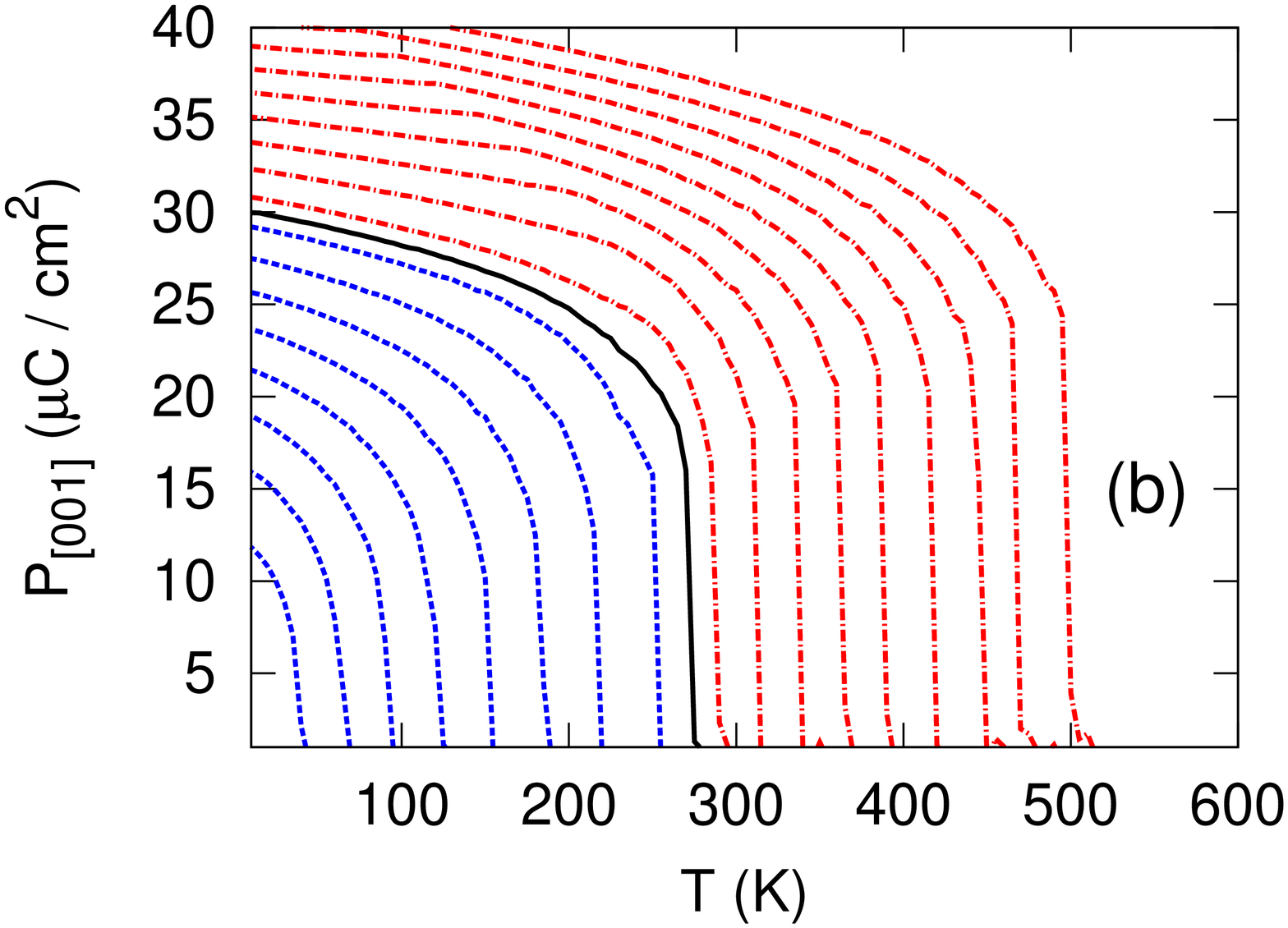}
\caption{(Color online) Calculated temperature-dependent polarization
  along (a) [110] and (b) [001] for different values of biaxial strain
  applied in the (001) plane (from $-$0.85\,\% to +1.75\,\% in steps
  of 0.1\,\%).  Solid (black) line: unstrained case ($\eta=0$); Dashed (blue)
  lines: tensile strain; Dash-dotted (red) lines: compressive
  strain.
\label{fig:P_strain}}
\end{figure}

Fig.~\ref{fig:P_strain} shows the calculated temperature dependence of
the total electric polarization $P$, separated into in-plane and
out-of-plane components, for different values of the epitaxial strain
($\eta_a=\eta_b$). It can be seen that for zero strain both in-plane
and out-of-plane components of $P$ appear at the same temperature,
while under compressive strain the appearance of the out-of-plane
(in-plane) components are shifted to higher (lower) temperatures, and
vice versa under tensile strain. This is in agreement with
Ref.~\onlinecite{Dieguez,Madhura}. However, we also observe that the
temperature dependence of the in-plane polarization under tensile
strain exhibits an unusual ``kink-like'' feature at a
(strain-dependent) temperature $T_S$, somewhat below the overall
critical temperature $T_C$.

\begin{figure*}
\centering
\includegraphics[width=0.6\textwidth,clip,trim=0cm  0cm 0cm 0cm]{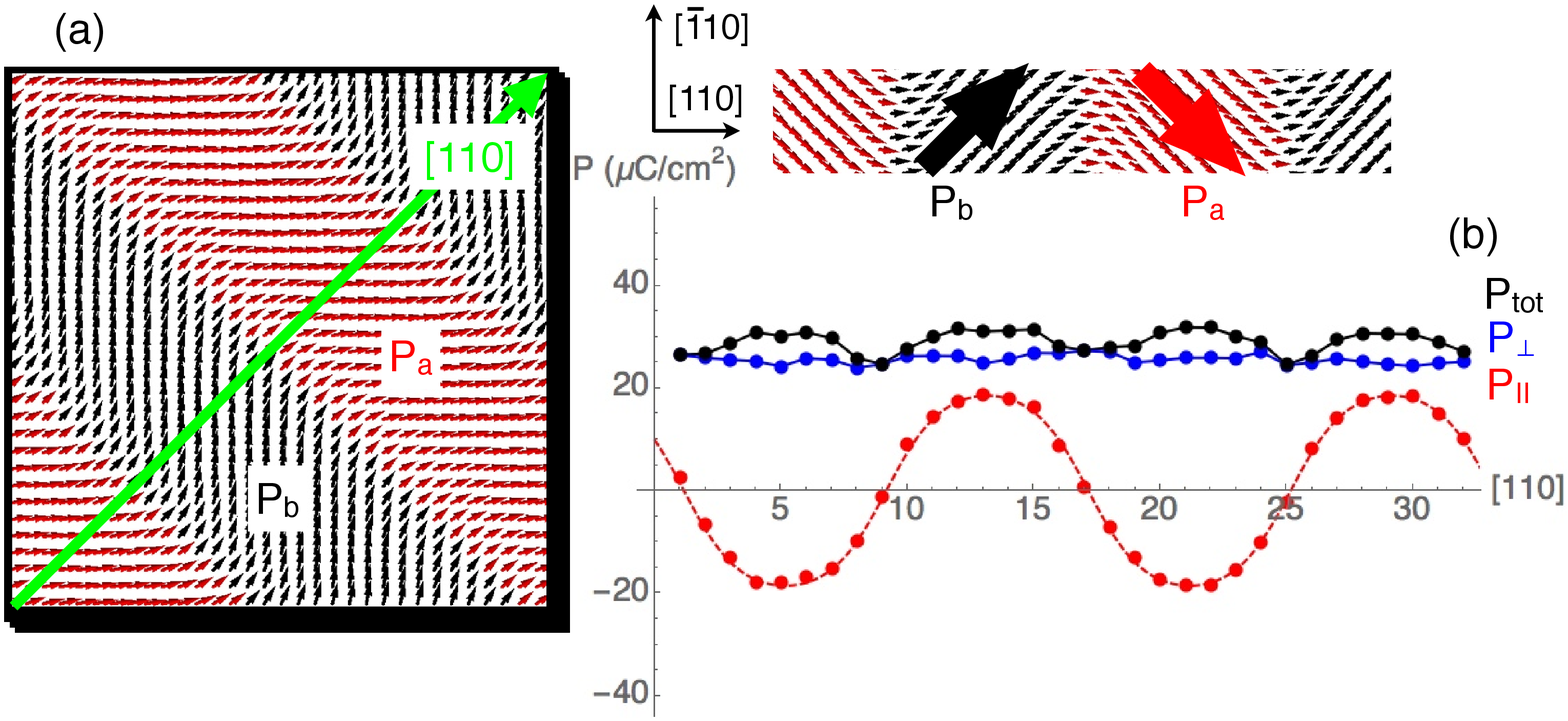}
\includegraphics[width=0.35\textwidth,clip,trim=1cm 2cm 0cm 0cm]{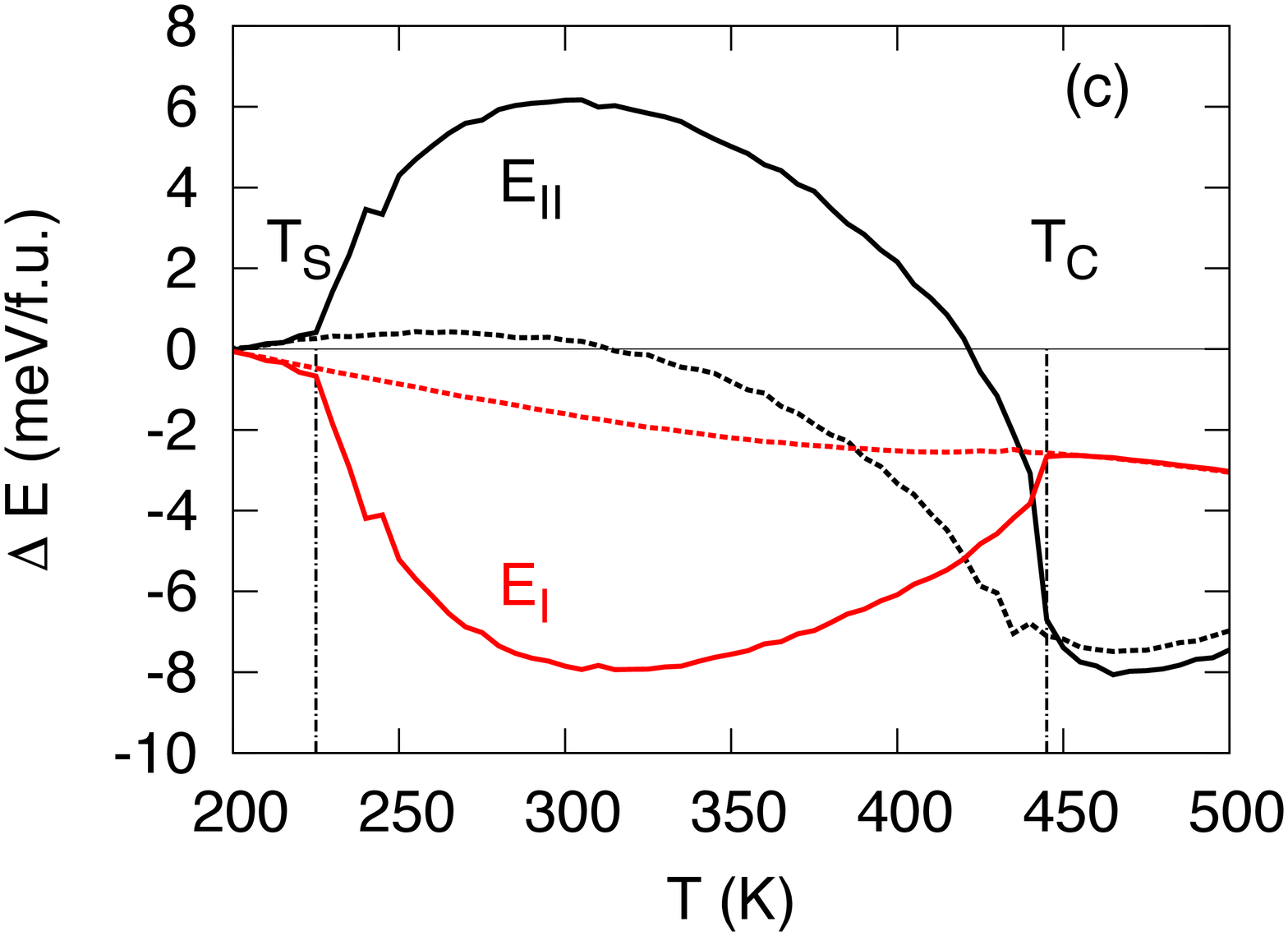}
\caption{(Color online) (a) Domain structure under 0.75\% tensile
  epitaxial strain for $T_C > T=400\,\mathrm{K} > T_S$. Local dipole
  moments have been averaged along $c$ and over 20 snapshots taken
  with a time distance of 0.2\,ps. 
(b) Cross section of the polarization profile along $[110]$. Black:
  total polarization; Blue: P$_{\perp}$; Red: P$_{||}$, see text. Dashed line:
  fit to the domain profile given by Eq.~\eqref{eq:tanh}. (c)
  Temperature dependence of selected energy contributions for
  $\eta=0.75$\,\%, normalized to 1 formula unit (f.u.) of BTO and taking the corresponding energy at 200\,K as reference. Solid
  lines: case with multi-domain states (32$\times$32$\times$32 cell);
  dashed lines: mono-domain case (16$\times$16$\times$16 cell). $E_{\text{I}}$:
  coupling between soft mode and local strain; $E_{\text{II}}$: local soft
  mode energy and inhomogeneous part of the elastic energy. Vertical
  lines indicate $T_C$ and $T_S$.
\label{fig:domains}}
\end{figure*}

Further inspection of the local soft mode configurations reveals that
in the temperature range between $T_C$ and $T_S$ the system exhibits a
multi-domain state, with average in-plane polarization along $[110]$,
but local in-plane polarization along $[100]$ and $[010]$, separated
by 90$^\circ$ domain walls parallel to (110) (see
Fig.~\ref{fig:domains}(a)). Below $T_S$, the system adopts a
mono-domain state with both local and global in-plane polarization
along $[110]$.
The formation of a multi-domain state under tensile strain allows the
material to essentially recover the same phase sequence as in the free
(unclamped) case, at least on a local level. At temperatures
immediately below $T_C$, the tetragonal ferroelectric phase with
polarization along $\langle 100 \rangle$ has the lowest free energy in
the unclamped case.\cite{Kumar} However, a corresponding mono-domain
state is strongly disfavored by the elastic boundary conditions
introduced through the epitaxial constraint. Thus, by forming the
observed multi-domain state, the system can lower its overall free
energy by achieving $\langle 100 \rangle$ polarization locally, at the
cost of introducing energetically unfavorable domain walls. Thereby,
the epitaxial constraint $\eta_a=\eta_b > 0$ promotes in-plane
polarization with an equal volume fraction of $[100]$ and $[010]$
domains. However, for cell sizes below 20$\times$20$\times$20, the
energy penalty for forming domain walls exceeds the gain in free
energy within the $\langle 100 \rangle$-polarized domains, and
consequently the multi-domain phase has been overlooked in previous
{\it ab initio} simulations employing smaller simulation cells.
Under further cooling, the $\langle 110 \rangle$-polarized phase
becomes more favorable, cf. Ref.~\onlinecite{Kumar} for the unclamped
material, and thus a transition into a corresponding mono-domain
state, which is also compatible with the epitaxial constraint under
tensile strain, occurs at $T_S$.

For local $\langle 100 \rangle$
polarization, only 180$^{\circ}$ domain walls and 90$^{\circ}$ walls
parallel to \{110\} are possible by symmetry.\cite{Marton} The latter type
is indeed observed in our simulations (see Fig.~\ref{fig:domains}(a)),
whereas 180$^{\circ}$ domain walls do not relax any elastic energy and
are thus not favorable in the present case. The polarization across
the 90$^\circ$ domain wall can be separated into the nearly constant
polarization perpendicular to the wall ($P_{\perp}$) and the
polarization parallel to the wall ($P_{\parallel}$), which
approximately follows a $\tanh$-profile:\cite{Kleemann,Marton}
\begin{equation}
P_{\parallel}(x) = P_{0,\parallel}\tanh\left[\frac{x_N}{4\pi d_{DW}}\sin
  \left(\frac{4\pi}{x_N} (x-x_0)\right)\right] \ ,
\label{eq:tanh}
\end{equation}
with $d_{DW}$ half of the domain wall width, $x_0$ center of one wall, $P_{0,\parallel}$, polarization in the domain center,
$x_N$ width of the fit region, and 4 the number of domains in the
simulation cell (a minimum of 4 domains is necessary to match the
domain profile at the periodic boundaries of the simulation cell). For
$T=400$\,K and 0.75\,\% tensile strain, the fit shown in
Fig.~\ref{fig:domains}\,(b) yields a domain wall width of about 5\,nm,
in good agreement with literature.\cite{Kleemann,Marton}

A clear signature of the multi-domain state in the temperature range
between $T_C$ and $T_S$ can also be seen in various energy
contributions, see Fig.~\ref{fig:domains}(c). The modulation of
$P_{\parallel}$ and the elastic mismatch at the domain walls induce an
energy penalty in the local mode self energy and the inhomogeneous
part of the elastic energy, whereas the coupling energy between local
strain and soft mode is reduced if the mono-domain state with $P$
along $\langle 110 \rangle$ breaks up into multiple domains with local
polarization along $\langle 100 \rangle$.
We use these anomalies in the different energy contributions to
identify $T_C$ and $T_S$ as function of strain and temperature. If no
clear jump in energy is visible at $T_C$ (due to the continuous
character of this transition under strain),\cite{pertsev2,Kumar} we
instead use the lowest temperature with $P=0$, which, however, gives rise to a large uncertainty of about 20~K.

\begin{figure*}
\centerline{\includegraphics[width=0.44\textwidth,clip,trim=1cm 3cm  10cm .5cm]{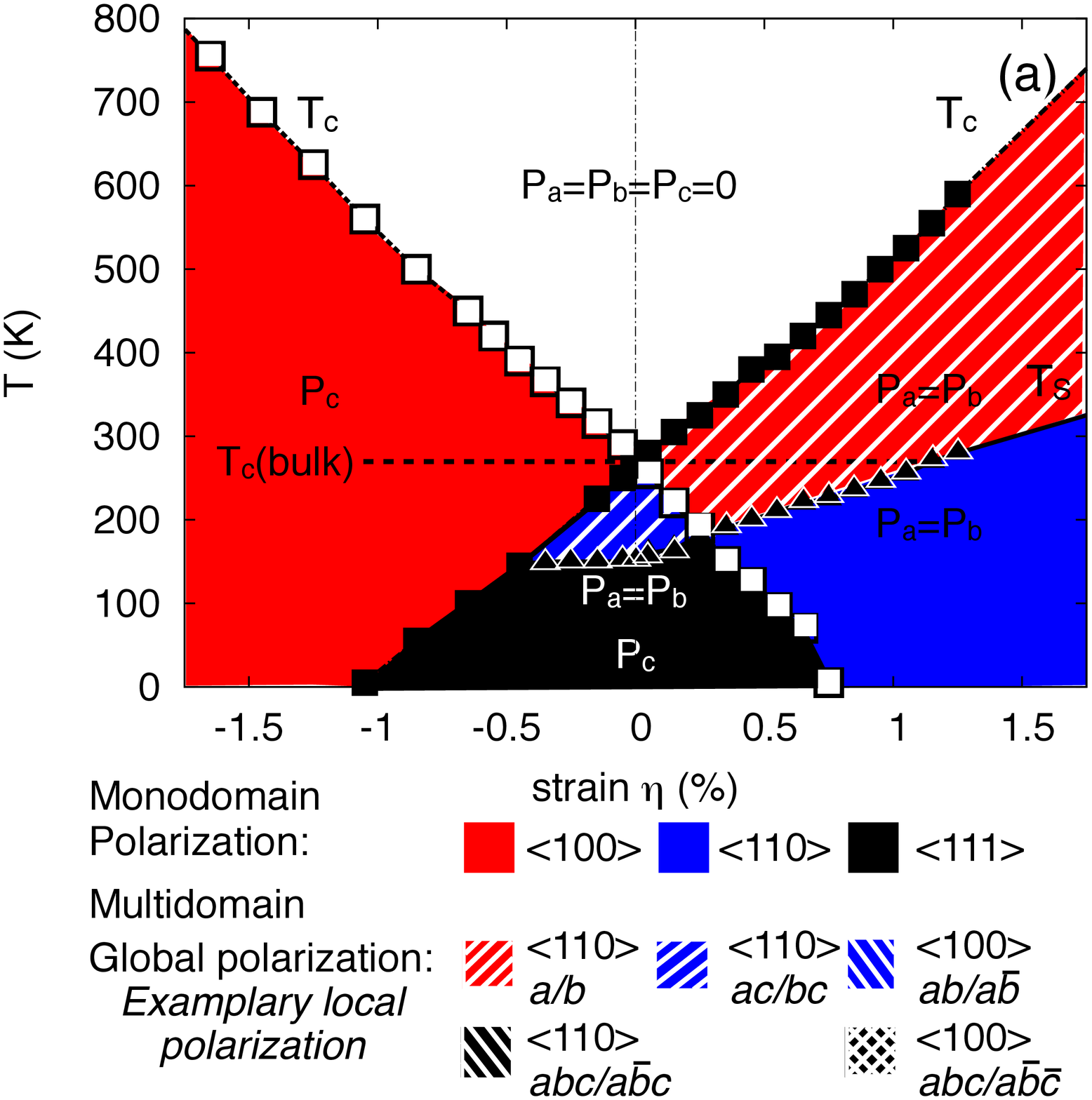}
  \includegraphics[width=0.63\textwidth,clip,trim= 0.1cm 1.5cm 2cm 2cm]{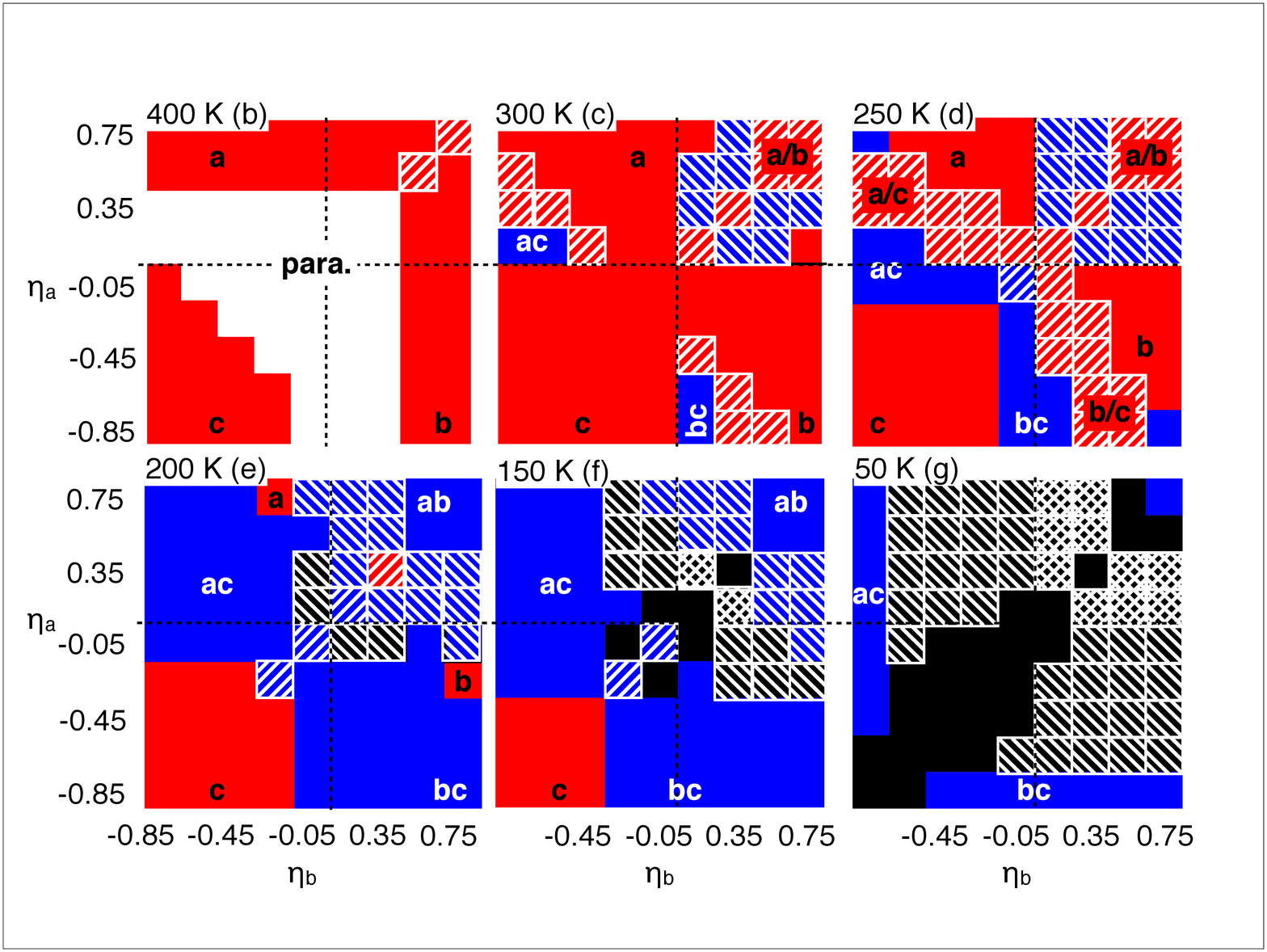}}
\caption{(Color online) Phase diagram of BaTiO$_3$ under biaxial
  strain. Colors/grayscale encode the paraelectric (white) and
  ferroelectric phases with local polarization along $\langle 100
  \rangle$ (red/light gray), $\langle 110 \rangle$ (blue/dark gray),
  and $\langle 111 \rangle$ (black). Multi-domain configurations are indicated
  by striped regions as specified in the legend. (a) ``Uniform''
  biaxial strain ($\eta_a = \eta_b$). Open (filled) squares indicate
  out-of-plane (in-plane) $T_C$, triangles indicate $T_S$. The
  horizontal dashed line indicates $T_C$ for the free bulk material.
  (b)-(g) General biaxial strain with $\eta_a \neq \eta_b$ for
  different temperatures.}
\label{fig:phase} 
\end{figure*}

Fig.~\ref{fig:phase}~(a) illustrates the so-obtained phase diagram of
BTO under uniform ($\eta_a=\eta_b$) biaxial strain. In qualitative
agreement with previous work,\cite{Dieguez,Madhura} we find two
transition lines, $T_C$, corresponding to the appearance of in-plane
and out-of-plane polarization, respectively, which cross for zero
strain at the transition temperature of the free bulk material. An
additional transition line, $T_S$, separating multi-domain and
single-domain phases is observed, extending from the tensile strain
region into the region of small compressive strain. We note that $T_S$
in Fig.~\ref{fig:phase}(a) is still somewhat dependent on the used
cell size (e.g. for $\eta=0.75$\%, $T_S$ is reduced by 35\,K
when using a 92$\times$92$\times$92 cell, while the the domain wall width is fully converged for a 48$\times$48$\times$48 cell).\\
Below both in-plane and out-of plane $T_C$, we obtain a multi-domain region where the in-plane components of the
polarization exhibit the same domain patterns as in the tensile strain
region (see e.g. Fig.~\ref{fig:domains}(a)), but with an additional
uniform polarization component along $c$, resulting in 60$^{\circ}$
domain walls parallel to \{110\}. 
Thus, in this region 
the local and global polarizations are along $\langle 110 \rangle$ and
$\langle 111 \rangle$, respectively. 
We note that down to $\eta=-0.45$\%, the imposed strain acts as tensile strain relative to the shorter lattice constant of the tetragonal phase in the free material. 
For stronger compressive strain, $T_S$ merges with the out-of-plane $T_C$ and
no multi-domain states are found. 

The case of general biaxial strain (with $\eta_a\not=\eta_b$) is
illustrated in Fig.~\ref{fig:phase}~(b)-(g). Under cooling, first a
local polarization along $\langle 100 \rangle$ (light, red regions)
appears in most regions of the phase diagram, with $P$ pointing
towards the longest lattice direction ($a$ or $b$ for $\eta_a$ or
$\eta_b >0$, $c$ for compressive strain). The corresponding
out-of-plane $T_C$ decreases linearly with $\eta_a$+$\eta_b$, while
the corresponding in-plane $T_C$ increases linearly with the strain
along the polarization direction, but is rather insensitive to the
strain in the perpendicular direction. If two directions are under
tensile strain, e.g.\ $b$ and $c$ for $\eta_a=-0.85$\% and
$\eta_b=0.35$\%, a multi-domain state with local polarization along
$\langle 100 \rangle$ and 90$^{\circ}$ domain walls is more favorable
compared to a mono-domain state with polarization along $\langle 110
\rangle$. Thus, both mono- or multi-domain phases with local
polarization along $\langle 100 \rangle$ are found below T$_C$.

Under further cooling, local polarization along $\langle 110 \rangle$
(darker, blue) and then along $\langle 111 \rangle$ (black) becomes
more favorable, analogously to the case of the free material. In
general, the tendency for polarization along a certain direction
increases, if the corresponding lattice constants in the clamped case
are larger compared to their cubic, tetragonal, or orthorhombic
counterparts in the free material.  Thus, the transition into the
$\langle 110 \rangle$ phases first sets in when both $\eta_a$ and
$\eta_b$ are positive and for a combination of large compressive and
weak tensile in-plane strains (leading to an elongation along
$c$). Again, multi-domain states occur for a broad range of strains
and temperatures. For local $\langle 110 \rangle$ polarization,
90$^{\circ}$, 60$^{\circ}$, and 120$^{\circ}$ domain walls are in
principle possible.\cite{Marton} As discussed above, only $60^{\circ}$
domain walls parallel to \{110\} with local
$ac$ and $bc$ polarization are observed for $\eta_a=\eta_b$. For $\eta_a\not=\eta_b$, we find also $90^{\circ}$
domain walls parallel to \{100\} with local
$ab$/$\bar{a}b$ or $ab$/$a\bar{b}$ polarization and vanishing average
global polarization along the shorter clamped lattice constant.
For local $\langle 111 \rangle$ polarization, no multi-domain states
are found for $\eta_a=\eta_b$. For $\eta_a\not=\eta_b$, 109$^{\circ}$
domain walls parallel to (010) and 71$^{\circ}$
walls parallel to (101) are possible,\cite{Marton} and both types are found in our
calculations. 71$^\circ$ walls occur for a combination of compressive
and tensile strain and the corresponding local $abc$/$\bar{a}bc$ or
$abc$/$a\bar{b}c$ polarization results in a vanishing global
polarization along the compressed lattice constant. 109$^\circ$ walls
with local $abc/a\bar{b}\bar{c}$ or $abc$/$\bar{a}b\bar{c}$ polarization, with neither 
 net polarization along the shorter clamped lattice constant nor along $c$, are found
for non-uniform tensile in-plane strain.

In summary, we have shown, using {\it ab initio}-based molecular
dynamics simulations, that the phase diagram of BTO under biaxial
strain shows a variety of multi-domain regions. The formation of
domains allows the system to recover the same sequence of para- and
ferroelectric phases as in the unstrained bulk material (at least for
most combinations of $\eta_a$ and $\eta_b$), while simultaneously
fulfilling all epitaxial constraints. Such multi-domain states have
not been found in previous {\it ab initio}-based studies of strained
BTO, due to restrictions in the size of the simulation cells. However,
similar multi-domain configurations,  have been observed in previous empirical phase-field
simulations.\cite{Chen:2008} Our results thus consolidate to some
extent the empirical with the {\it ab initio}-based simulations, and
provide new insights for future studies and the better interpretation
of experimental data in strained BTO films.
\section*{Acknowledgment}
We acknowledge financial support by the Deutsche
Forschungsgemeinschaft (SPP 1599) and the Swiss National Science
Foundation, and thank the Center for Computational Science and
Simulation at the University of Duisburg-Essen for computer time.

\bibliography{anna}

\begin{thebibliography}{30}%
\makeatletter
\providecommand \@ifxundefined [1]{%
 \@ifx{#1\undefined}
}%
\providecommand \@ifnum [1]{%
 \ifnum #1\expandafter \@firstoftwo
 \else \expandafter \@secondoftwo
 \fi
}%
\providecommand \@ifx [1]{%
 \ifx #1\expandafter \@firstoftwo
 \else \expandafter \@secondoftwo
 \fi
}%
\providecommand \natexlab [1]{#1}%
\providecommand \enquote  [1]{``#1''}%
\providecommand \bibnamefont  [1]{#1}%
\providecommand \bibfnamefont [1]{#1}%
\providecommand \citenamefont [1]{#1}%
\providecommand \href@noop [0]{\@secondoftwo}%
\providecommand \href [0]{\begingroup \@sanitize@url \@href}%
\providecommand \@href[1]{\@@startlink{#1}\@@href}%
\providecommand \@@href[1]{\endgroup#1\@@endlink}%
\providecommand \@sanitize@url [0]{\catcode `\\12\catcode `\$12\catcode
  `\&12\catcode `\#12\catcode `\^12\catcode `\_12\catcode `\%12\relax}%
\providecommand \@@startlink[1]{}%
\providecommand \@@endlink[0]{}%
\providecommand \url  [0]{\begingroup\@sanitize@url \@url }%
\providecommand \@url [1]{\endgroup\@href {#1}{\urlprefix }}%
\providecommand \urlprefix  [0]{URL }%
\providecommand \Eprint [0]{\href }%
\providecommand \doibase [0]{http://dx.doi.org/}%
\providecommand \selectlanguage [0]{\@gobble}%
\providecommand \bibinfo  [0]{\@secondoftwo}%
\providecommand \bibfield  [0]{\@secondoftwo}%
\providecommand \translation [1]{[#1]}%
\providecommand \BibitemOpen [0]{}%
\providecommand \bibitemStop [0]{}%
\providecommand \bibitemNoStop [0]{.\EOS\space}%
\providecommand \EOS [0]{\spacefactor3000\relax}%
\providecommand \BibitemShut  [1]{\csname bibitem#1\endcsname}%
\let\auto@bib@innerbib\@empty
\bibitem [{\citenamefont {Schlom}\ \emph {et~al.}(2007)\citenamefont {Schlom},
  \citenamefont {Chen}, \citenamefont {Eom}, \citenamefont {Rabe},
  \citenamefont {Streiffer},\ and\ \citenamefont {Triscone}}]{Schlom}%
  \BibitemOpen
  \bibfield  {author} {\bibinfo {author} {\bibfnamefont {D.~G.}\ \bibnamefont
  {Schlom}}, \bibinfo {author} {\bibfnamefont {L.-Q.}\ \bibnamefont {Chen}},
  \bibinfo {author} {\bibfnamefont {C.-B.}\ \bibnamefont {Eom}}, \bibinfo
  {author} {\bibfnamefont {K.~M.}\ \bibnamefont {Rabe}}, \bibinfo {author}
  {\bibfnamefont {S.~K.}\ \bibnamefont {Streiffer}}, \ and\ \bibinfo {author}
  {\bibfnamefont {J.-M.}\ \bibnamefont {Triscone}},\ }\href@noop {} {\bibfield
  {journal} {\bibinfo  {journal} {Ann. Rev. Mater. Research}\ }\textbf
  {\bibinfo {volume} {37}},\ \bibinfo {pages} {586} (\bibinfo {year}
  {2007})}\BibitemShut {NoStop}%
\bibitem [{\citenamefont {Schlom}\ \emph {et~al.}(2008)\citenamefont {Schlom},
  \citenamefont {Chen}, \citenamefont {Pan}, \citenamefont {Schmehl},\ and\
  \citenamefont {Zurbuchen}}]{Schlom_et_al:2008}%
  \BibitemOpen
  \bibfield  {author} {\bibinfo {author} {\bibfnamefont {D.~G.}\ \bibnamefont
  {Schlom}}, \bibinfo {author} {\bibfnamefont {L.-Q.}\ \bibnamefont {Chen}},
  \bibinfo {author} {\bibfnamefont {X.}~\bibnamefont {Pan}}, \bibinfo {author}
  {\bibfnamefont {A.}~\bibnamefont {Schmehl}}, \ and\ \bibinfo {author}
  {\bibfnamefont {M.~A.}\ \bibnamefont {Zurbuchen}},\ }\href {\doibase
  10.1111/j.1551-2916.2008.02556.x} {\bibfield  {journal} {\bibinfo  {journal}
  {Journal of the American Ceramic Society}\ }\textbf {\bibinfo {volume}
  {91}},\ \bibinfo {pages} {2429} (\bibinfo {year} {2008})}\BibitemShut
  {NoStop}%
\bibitem [{\citenamefont {Biegalski}\ \emph {et~al.}(2010)\citenamefont
  {Biegalski}, \citenamefont {D{\"o}rr}, \citenamefont {Kim},\ and\
  \citenamefont {Christen}}]{Biegalski_et_al:2010}%
  \BibitemOpen
  \bibfield  {author} {\bibinfo {author} {\bibfnamefont {M.~D.}\ \bibnamefont
  {Biegalski}}, \bibinfo {author} {\bibfnamefont {K.}~\bibnamefont {D{\"o}rr}},
  \bibinfo {author} {\bibfnamefont {D.~H.}\ \bibnamefont {Kim}}, \ and\
  \bibinfo {author} {\bibfnamefont {H.~M.}\ \bibnamefont {Christen}},\ }\href
  {\doibase 10.1063/1.1374323} {\bibfield  {journal} {\bibinfo  {journal}
  {Applied Physics Letters}\ }\textbf {\bibinfo {volume} {96}},\ \bibinfo
  {pages} {151905} (\bibinfo {year} {2010})}\BibitemShut {NoStop}%
\bibitem [{\citenamefont {Toshio}\ \emph {et~al.}(1981)\citenamefont {Toshio},
  \citenamefont {Hellwege}, \citenamefont {Landolt}, \citenamefont
  {B{\"o}rnstein},\ and\ \citenamefont {Madelung}}]{Toshio}%
  \BibitemOpen
  \bibfield  {author} {\bibinfo {author} {\bibfnamefont {T.}~\bibnamefont
  {Toshio}}, \bibinfo {author} {\bibfnamefont {K.-H.}\ \bibnamefont
  {Hellwege}}, \bibinfo {author} {\bibfnamefont {H.}~\bibnamefont {Landolt}},
  \bibinfo {author} {\bibfnamefont {R.}~\bibnamefont {B{\"o}rnstein}}, \ and\
  \bibinfo {author} {\bibfnamefont {O.}~\bibnamefont {Madelung}},\ }in\
  \href@noop {} {\emph {\bibinfo {booktitle} {Landolt-Bornstein Numerical Data
  and Functional Reltionships in Science and Technology}}},\ \bibinfo {series}
  {Group III}, Vol.~\bibinfo {volume} {3}\ (\bibinfo  {publisher} {Springer,
  Berlin},\ \bibinfo {year} {1981})\BibitemShut {NoStop}%
\bibitem [{\citenamefont {Pertsev}, \citenamefont {Zembilgotov},\ and\
  \citenamefont {Tagantsev}(1998)}]{pertsev2}%
  \BibitemOpen
  \bibfield  {author} {\bibinfo {author} {\bibfnamefont {N.~A.}\ \bibnamefont
  {Pertsev}}, \bibinfo {author} {\bibfnamefont {A.~G.}\ \bibnamefont
  {Zembilgotov}}, \ and\ \bibinfo {author} {\bibfnamefont {A.~K.}\ \bibnamefont
  {Tagantsev}},\ }\href@noop {} {\bibfield  {journal} {\bibinfo  {journal}
  {Phys. Rev. Lett.}\ }\textbf {\bibinfo {volume} {80}},\ \bibinfo {pages}
  {1988} (\bibinfo {year} {1998})}\BibitemShut {NoStop}%
\bibitem [{\citenamefont {Pertsev}\ and\ \citenamefont
  {Koukhar}(2000)}]{pertsev4}%
  \BibitemOpen
  \bibfield  {author} {\bibinfo {author} {\bibfnamefont {N.~A.}\ \bibnamefont
  {Pertsev}}\ and\ \bibinfo {author} {\bibfnamefont {V.~G.}\ \bibnamefont
  {Koukhar}},\ }\href@noop {} {\bibfield  {journal} {\bibinfo  {journal} {Phys.
  Rev. Lett.}\ }\textbf {\bibinfo {volume} {84}},\ \bibinfo {pages} {3722}
  (\bibinfo {year} {2000})}\BibitemShut {NoStop}%
\bibitem [{\citenamefont {Pertsev}\ \emph {et~al.}(2001)\citenamefont
  {Pertsev}, \citenamefont {Koukhar}, \citenamefont {Waser},\ and\
  \citenamefont {Hoffmann}}]{pertsev3}%
  \BibitemOpen
  \bibfield  {author} {\bibinfo {author} {\bibfnamefont {N.~A.}\ \bibnamefont
  {Pertsev}}, \bibinfo {author} {\bibfnamefont {V.~G.}\ \bibnamefont
  {Koukhar}}, \bibinfo {author} {\bibfnamefont {R.}~\bibnamefont {Waser}}, \
  and\ \bibinfo {author} {\bibfnamefont {S.}~\bibnamefont {Hoffmann}},\
  }\href@noop {} {\bibfield  {journal} {\bibinfo  {journal} {Integrated
  ferroelectrics}\ }\textbf {\bibinfo {volume} {32}},\ \bibinfo {pages} {235}
  (\bibinfo {year} {2001})}\BibitemShut {NoStop}%
\bibitem [{\citenamefont {Pertsev}, \citenamefont {Zembilgotov},\ and\
  \citenamefont {Tagantsev}(2011)}]{pertsev}%
  \BibitemOpen
  \bibfield  {author} {\bibinfo {author} {\bibfnamefont {N.~A.}\ \bibnamefont
  {Pertsev}}, \bibinfo {author} {\bibfnamefont {A.~G.}\ \bibnamefont
  {Zembilgotov}}, \ and\ \bibinfo {author} {\bibfnamefont {A.~K.}\ \bibnamefont
  {Tagantsev}},\ }\href@noop {} {\bibfield  {journal} {\bibinfo  {journal}
  {Ferroelectrics}\ }\textbf {\bibinfo {volume} {223}},\ \bibinfo {pages} {79}
  (\bibinfo {year} {2011})}\BibitemShut {NoStop}%
\bibitem [{\citenamefont {Choi}\ \emph {et~al.}(2004)\citenamefont {Choi},
  \citenamefont {Biegalski}, \citenamefont {Li}, \citenamefont {Sharan},
  \citenamefont {Schubert}, \citenamefont {Uecker}, \citenamefont {Reiche},
  \citenamefont {Chen}, \citenamefont {Pan}, \citenamefont {Gopalan},
  \citenamefont {Chen}, \citenamefont {Schlom},\ and\ \citenamefont
  {Eom}}]{choi}%
  \BibitemOpen
  \bibfield  {author} {\bibinfo {author} {\bibfnamefont {K.~J.}\ \bibnamefont
  {Choi}}, \bibinfo {author} {\bibfnamefont {M.}~\bibnamefont {Biegalski}},
  \bibinfo {author} {\bibfnamefont {Y.}~\bibnamefont {Li}}, \bibinfo {author}
  {\bibfnamefont {A.}~\bibnamefont {Sharan}}, \bibinfo {author} {\bibfnamefont
  {J.}~\bibnamefont {Schubert}}, \bibinfo {author} {\bibfnamefont
  {R.}~\bibnamefont {Uecker}}, \bibinfo {author} {\bibfnamefont
  {P.}~\bibnamefont {Reiche}}, \bibinfo {author} {\bibfnamefont {Y.~B.}\
  \bibnamefont {Chen}}, \bibinfo {author} {\bibfnamefont {X.~Q.}\ \bibnamefont
  {Pan}}, \bibinfo {author} {\bibfnamefont {V.}~\bibnamefont {Gopalan}},
  \bibinfo {author} {\bibfnamefont {L.-Q.}\ \bibnamefont {Chen}}, \bibinfo
  {author} {\bibfnamefont {D.~G.}\ \bibnamefont {Schlom}}, \ and\ \bibinfo
  {author} {\bibfnamefont {C.}~\bibnamefont {Eom}},\ }\href@noop {} {\bibfield
  {journal} {\bibinfo  {journal} {Science}\ }\textbf {\bibinfo {volume}
  {306}},\ \bibinfo {pages} {1005} (\bibinfo {year} {2004})}\BibitemShut
  {NoStop}%
\bibitem [{\citenamefont {Li}\ \emph {et~al.}(2001)\citenamefont {Li},
  \citenamefont {Hu}, \citenamefont {Liu},\ and\ \citenamefont
  {Chen}}]{Li_et_al:2001}%
  \BibitemOpen
  \bibfield  {author} {\bibinfo {author} {\bibfnamefont {Y.~L.}\ \bibnamefont
  {Li}}, \bibinfo {author} {\bibfnamefont {S.~Y.}\ \bibnamefont {Hu}}, \bibinfo
  {author} {\bibfnamefont {Z.~K.}\ \bibnamefont {Liu}}, \ and\ \bibinfo
  {author} {\bibfnamefont {L.~Q.}\ \bibnamefont {Chen}},\ }\href {\doibase
  10.1063/1.1377855} {\bibfield  {journal} {\bibinfo  {journal} {Applied
  Physics Letters}\ }\textbf {\bibinfo {volume} {78}},\ \bibinfo {pages} {3878}
  (\bibinfo {year} {2001})}\BibitemShut {NoStop}%
\bibitem [{\citenamefont {Li}\ and\ \citenamefont {Chen}(2006)}]{Li2}%
  \BibitemOpen
  \bibfield  {author} {\bibinfo {author} {\bibfnamefont {Y.~L.}\ \bibnamefont
  {Li}}\ and\ \bibinfo {author} {\bibfnamefont {L.~Q.}\ \bibnamefont {Chen}},\
  }\href@noop {} {\bibfield  {journal} {\bibinfo  {journal} {Appl. Phys.
  Lett.}\ }\textbf {\bibinfo {volume} {88}},\ \bibinfo {pages} {072905}
  (\bibinfo {year} {2006})}\BibitemShut {NoStop}%
\bibitem [{\citenamefont {Chen}(2008)}]{Chen:2008}%
  \BibitemOpen
  \bibfield  {author} {\bibinfo {author} {\bibfnamefont {L.-Q.}\ \bibnamefont
  {Chen}},\ }\href {\doibase 10.1111/j.1551-2916.2008.02413.x} {\bibfield
  {journal} {\bibinfo  {journal} {Journal of the American Ceramic Society}\
  }\textbf {\bibinfo {volume} {91}},\ \bibinfo {pages} {1835} (\bibinfo {year}
  {2008})}\BibitemShut {NoStop}%
\bibitem [{\citenamefont {Di{\'{e}}guez}\ \emph {et~al.}(2004)\citenamefont
  {Di{\'{e}}guez}, \citenamefont {Tinte}, \citenamefont {Antons}, \citenamefont
  {Bungaro}, \citenamefont {Neaton}, \citenamefont {Rabe},\ and\ \citenamefont
  {Vanderbilt}}]{Dieguez}%
  \BibitemOpen
  \bibfield  {author} {\bibinfo {author} {\bibfnamefont {O.}~\bibnamefont
  {Di{\'{e}}guez}}, \bibinfo {author} {\bibfnamefont {S.}~\bibnamefont
  {Tinte}}, \bibinfo {author} {\bibfnamefont {A.}~\bibnamefont {Antons}},
  \bibinfo {author} {\bibfnamefont {C.}~\bibnamefont {Bungaro}}, \bibinfo
  {author} {\bibfnamefont {J.~B.}\ \bibnamefont {Neaton}}, \bibinfo {author}
  {\bibfnamefont {K.~M.}\ \bibnamefont {Rabe}}, \ and\ \bibinfo {author}
  {\bibfnamefont {D.}~\bibnamefont {Vanderbilt}},\ }\href@noop {} {\bibfield
  {journal} {\bibinfo  {journal} {Phys. Rev. B}\ }\textbf {\bibinfo {volume}
  {69}},\ \bibinfo {pages} {212101} (\bibinfo {year} {2004})}\BibitemShut
  {NoStop}%
\bibitem [{\citenamefont {Lai}\ \emph {et~al.}(2005)\citenamefont {Lai},
  \citenamefont {Kornev}, \citenamefont {Bellaiche},\ and\ \citenamefont
  {Salamo}}]{Lai_et_al:2004}%
  \BibitemOpen
  \bibfield  {author} {\bibinfo {author} {\bibfnamefont {B.-K.}\ \bibnamefont
  {Lai}}, \bibinfo {author} {\bibfnamefont {I.}~\bibnamefont {Kornev}},
  \bibinfo {author} {\bibfnamefont {L.}~\bibnamefont {Bellaiche}}, \ and\
  \bibinfo {author} {\bibfnamefont {G.}~\bibnamefont {Salamo}},\ }\href@noop {}
  {\bibfield  {journal} {\bibinfo  {journal} {Appl. Phys. Lett.}\ }\textbf
  {\bibinfo {volume} {86}},\ \bibinfo {pages} {132904} (\bibinfo {year}
  {2005})}\BibitemShut {NoStop}%
\bibitem [{\citenamefont {Paul}\ \emph {et~al.}(2007)\citenamefont {Paul},
  \citenamefont {Nishimatsu}, \citenamefont {Kawazoe},\ and\ \citenamefont
  {Waghmare}}]{Feram2}%
  \BibitemOpen
  \bibfield  {author} {\bibinfo {author} {\bibfnamefont {J.}~\bibnamefont
  {Paul}}, \bibinfo {author} {\bibfnamefont {T.}~\bibnamefont {Nishimatsu}},
  \bibinfo {author} {\bibfnamefont {Y.}~\bibnamefont {Kawazoe}}, \ and\
  \bibinfo {author} {\bibfnamefont {U.~V.}\ \bibnamefont {Waghmare}},\
  }\href@noop {} {\bibfield  {journal} {\bibinfo  {journal} {Phys. Rev. Lett.}\
  }\textbf {\bibinfo {volume} {99}},\ \bibinfo {pages} {077601} (\bibinfo
  {year} {2007})}\BibitemShut {NoStop}%
\bibitem [{\citenamefont {Marathe}\ and\ \citenamefont
  {Ederer}(2014)}]{Madhura}%
  \BibitemOpen
  \bibfield  {author} {\bibinfo {author} {\bibfnamefont {M.}~\bibnamefont
  {Marathe}}\ and\ \bibinfo {author} {\bibfnamefont {C.}~\bibnamefont
  {Ederer}},\ }\href@noop {} {\bibfield  {journal} {\bibinfo  {journal} {App.
  Phys. Lett.}\ }\textbf {\bibinfo {volume} {104}},\ \bibinfo {pages} {212902}
  (\bibinfo {year} {2014})}\BibitemShut {NoStop}%
\bibitem [{\citenamefont {Pompe}\ \emph {et~al.}(1993)\citenamefont {Pompe},
  \citenamefont {Gong}, \citenamefont {Suo},\ and\ \citenamefont
  {Speck}}]{Pompe}%
  \BibitemOpen
  \bibfield  {author} {\bibinfo {author} {\bibfnamefont {W.}~\bibnamefont
  {Pompe}}, \bibinfo {author} {\bibfnamefont {X.}~\bibnamefont {Gong}},
  \bibinfo {author} {\bibfnamefont {Z.}~\bibnamefont {Suo}}, \ and\ \bibinfo
  {author} {\bibfnamefont {J.~S.}\ \bibnamefont {Speck}},\ }\href@noop {}
  {\bibfield  {journal} {\bibinfo  {journal} {J. Appl. Phys.}\ }\textbf
  {\bibinfo {volume} {74}},\ \bibinfo {pages} {6012} (\bibinfo {year}
  {1993})}\BibitemShut {NoStop}%
\bibitem [{\citenamefont {Kouser}, \citenamefont {Nishimatsu},\ and\
  \citenamefont {Waghmare}(2013)}]{Kouser}%
  \BibitemOpen
  \bibfield  {author} {\bibinfo {author} {\bibfnamefont {S.}~\bibnamefont
  {Kouser}}, \bibinfo {author} {\bibfnamefont {T.}~\bibnamefont {Nishimatsu}},
  \ and\ \bibinfo {author} {\bibfnamefont {U.~V.}\ \bibnamefont {Waghmare}},\
  }\href@noop {} {\bibfield  {journal} {\bibinfo  {journal} {Phys. Rev. B}\
  }\textbf {\bibinfo {volume} {88}},\ \bibinfo {pages} {064102} (\bibinfo
  {year} {2013})}\BibitemShut {NoStop}%
\bibitem [{\citenamefont {Gr\"unebohm}, \citenamefont {Gruner},\ and\
  \citenamefont {Entel}(2012)}]{Kleemann}%
  \BibitemOpen
  \bibfield  {author} {\bibinfo {author} {\bibfnamefont {A.}~\bibnamefont
  {Gr\"unebohm}}, \bibinfo {author} {\bibfnamefont {M.~E.}\ \bibnamefont
  {Gruner}}, \ and\ \bibinfo {author} {\bibfnamefont {P.}~\bibnamefont
  {Entel}},\ }\href@noop {} {\bibfield  {journal} {\bibinfo  {journal}
  {Ferroelectrics}\ }\textbf {\bibinfo {volume} {426}},\ \bibinfo {pages} {21}
  (\bibinfo {year} {2012})}\BibitemShut {NoStop}%
\bibitem [{\citenamefont {Nishimatsu}\ \emph {et~al.}(2012)\citenamefont
  {Nishimatsu}, \citenamefont {Aoyagi}, \citenamefont {Konno}, \citenamefont
  {Kawazoe}, \citenamefont {Funkaubo}, \citenamefont {Kumar},\ and\
  \citenamefont {Waghmare}}]{Nishimatsu2}%
  \BibitemOpen
  \bibfield  {author} {\bibinfo {author} {\bibfnamefont {T.}~\bibnamefont
  {Nishimatsu}}, \bibinfo {author} {\bibfnamefont {K.}~\bibnamefont {Aoyagi}},
  \bibinfo {author} {\bibfnamefont {T.~J.}\ \bibnamefont {Konno}}, \bibinfo
  {author} {\bibfnamefont {Y.}~\bibnamefont {Kawazoe}}, \bibinfo {author}
  {\bibfnamefont {H.}~\bibnamefont {Funkaubo}}, \bibinfo {author}
  {\bibfnamefont {A.}~\bibnamefont {Kumar}}, \ and\ \bibinfo {author}
  {\bibfnamefont {U.~V.}\ \bibnamefont {Waghmare}},\ }\href@noop {} {\bibfield
  {journal} {\bibinfo  {journal} {J. Phys. Soc. Jpn.}\ }\textbf {\bibinfo
  {volume} {81}},\ \bibinfo {pages} {124702} (\bibinfo {year}
  {2012})}\BibitemShut {NoStop}%
\bibitem [{\citenamefont {Nishimatsu}\ \emph {et~al.}(2008)\citenamefont
  {Nishimatsu}, \citenamefont {Waghmare}, \citenamefont {Kawazoe},\ and\
  \citenamefont {Vanderbilt}}]{Feram1}%
  \BibitemOpen
  \bibfield  {author} {\bibinfo {author} {\bibfnamefont {T.}~\bibnamefont
  {Nishimatsu}}, \bibinfo {author} {\bibfnamefont {U.~V.}\ \bibnamefont
  {Waghmare}}, \bibinfo {author} {\bibfnamefont {Y.}~\bibnamefont {Kawazoe}}, \
  and\ \bibinfo {author} {\bibfnamefont {D.}~\bibnamefont {Vanderbilt}},\
  }\href@noop {} {\bibfield  {journal} {\bibinfo  {journal} {Phys. Rev. B}\
  }\textbf {\bibinfo {volume} {78}},\ \bibinfo {pages} {104104} (\bibinfo
  {year} {2008})}\BibitemShut {NoStop}%
\bibitem [{\citenamefont {Qiao}\ and\ \citenamefont {Bi}(2008)}]{Qiao}%
  \BibitemOpen
  \bibfield  {author} {\bibinfo {author} {\bibfnamefont {L.}~\bibnamefont
  {Qiao}}\ and\ \bibinfo {author} {\bibfnamefont {X.}~\bibnamefont {Bi}},\
  }\href@noop {} {\bibfield  {journal} {\bibinfo  {journal} {App. Phys. Lett.}\
  }\textbf {\bibinfo {volume} {92}},\ \bibinfo {pages} {062912} (\bibinfo
  {year} {2008})}\BibitemShut {NoStop}%
\bibitem [{\citenamefont {Misirlioglu}\ \emph {et~al.}(2006)\citenamefont
  {Misirlioglu}, \citenamefont {Alpay}, \citenamefont {He},\ and\ \citenamefont
  {Wells}}]{Misirlioglu}%
  \BibitemOpen
  \bibfield  {author} {\bibinfo {author} {\bibfnamefont {I.~B.}\ \bibnamefont
  {Misirlioglu}}, \bibinfo {author} {\bibfnamefont {S.~P.}\ \bibnamefont
  {Alpay}}, \bibinfo {author} {\bibfnamefont {F.}~\bibnamefont {He}}, \ and\
  \bibinfo {author} {\bibfnamefont {B.~O.}\ \bibnamefont {Wells}},\ }\href@noop
  {} {\bibfield  {journal} {\bibinfo  {journal} {J. Appl. Phys.}\ }\textbf
  {\bibinfo {volume} {99}},\ \bibinfo {pages} {104103} (\bibinfo {year}
  {2006})}\BibitemShut {NoStop}%
\bibitem [{\citenamefont {He}\ and\ \citenamefont {Wells}(2006)}]{He}%
  \BibitemOpen
  \bibfield  {author} {\bibinfo {author} {\bibfnamefont {F.}~\bibnamefont
  {He}}\ and\ \bibinfo {author} {\bibfnamefont {B.~O.}\ \bibnamefont {Wells}},\
  }\href@noop {} {\bibfield  {journal} {\bibinfo  {journal} {Applied Physics
  Letters}\ }\textbf {\bibinfo {volume} {88}},\ \bibinfo {eid} {152908}
  (\bibinfo {year} {2006})}\BibitemShut {NoStop}%
\bibitem [{\citenamefont {King-Smith}\ and\ \citenamefont
  {Vanderbilt}(1994)}]{King}%
  \BibitemOpen
  \bibfield  {author} {\bibinfo {author} {\bibfnamefont {R.~D.}\ \bibnamefont
  {King-Smith}}\ and\ \bibinfo {author} {\bibfnamefont {D.}~\bibnamefont
  {Vanderbilt}},\ }\href {\doibase 10.1103/PhysRevB.49.5828} {\bibfield
  {journal} {\bibinfo  {journal} {Phys. Rev. B}\ }\textbf {\bibinfo {volume}
  {49}},\ \bibinfo {pages} {5828} (\bibinfo {year} {1994})}\BibitemShut
  {NoStop}%
\bibitem [{\citenamefont {Zhong}, \citenamefont {Vanderbilt},\ and\
  \citenamefont {Rabe}(1995)}]{Zhong}%
  \BibitemOpen
  \bibfield  {author} {\bibinfo {author} {\bibfnamefont {W.}~\bibnamefont
  {Zhong}}, \bibinfo {author} {\bibfnamefont {D.}~\bibnamefont {Vanderbilt}}, \
  and\ \bibinfo {author} {\bibfnamefont {K.~M.}\ \bibnamefont {Rabe}},\ }\href
  {\doibase 10.1103/PhysRevB.52.6301} {\bibfield  {journal} {\bibinfo
  {journal} {Phys. Rev. B}\ }\textbf {\bibinfo {volume} {52}},\ \bibinfo
  {pages} {6301} (\bibinfo {year} {1995})}\BibitemShut {NoStop}%
\bibitem [{\citenamefont {Nishimatsu}\ \emph {et~al.}(2010)\citenamefont
  {Nishimatsu}, \citenamefont {Iwamoto}, \citenamefont {Kawazoe},\ and\
  \citenamefont {Waghmare}}]{Nishimatsu}%
  \BibitemOpen
  \bibfield  {author} {\bibinfo {author} {\bibfnamefont {T.}~\bibnamefont
  {Nishimatsu}}, \bibinfo {author} {\bibfnamefont {M.}~\bibnamefont {Iwamoto}},
  \bibinfo {author} {\bibfnamefont {Y.}~\bibnamefont {Kawazoe}}, \ and\
  \bibinfo {author} {\bibfnamefont {U.~V.}\ \bibnamefont {Waghmare}},\
  }\href@noop {} {\bibfield  {journal} {\bibinfo  {journal} {Phys Rev B}\
  }\textbf {\bibinfo {volume} {82}},\ \bibinfo {pages} {134106} (\bibinfo
  {year} {2010})}\BibitemShut {NoStop}%
\bibitem [{\citenamefont {Bond}, \citenamefont {Leimkuhler},\ and\
  \citenamefont {Laird}(1999)}]{Nose}%
  \BibitemOpen
  \bibfield  {author} {\bibinfo {author} {\bibfnamefont {S.~D.}\ \bibnamefont
  {Bond}}, \bibinfo {author} {\bibfnamefont {B.~J.}\ \bibnamefont
  {Leimkuhler}}, \ and\ \bibinfo {author} {\bibfnamefont {B.~B.}\ \bibnamefont
  {Laird}},\ }\href {\doibase http://dx.doi.org/10.1006/jcph.1998.6171}
  {\bibfield  {journal} {\bibinfo  {journal} {Journal of Computational
  Physics}\ }\textbf {\bibinfo {volume} {151}},\ \bibinfo {pages} {114 }
  (\bibinfo {year} {1999})}\BibitemShut {NoStop}%
\bibitem [{\citenamefont {Marton}, \citenamefont {Rychetsky},\ and\
  \citenamefont {Hlinka}(2010)}]{Marton}%
  \BibitemOpen
  \bibfield  {author} {\bibinfo {author} {\bibfnamefont {P.}~\bibnamefont
  {Marton}}, \bibinfo {author} {\bibfnamefont {I.}~\bibnamefont {Rychetsky}}, \
  and\ \bibinfo {author} {\bibfnamefont {J.}~\bibnamefont {Hlinka}},\ }\href
  {\doibase 10.1103/PhysRevB.81.144125} {\bibfield  {journal} {\bibinfo
  {journal} {Phys. Rev. B}\ }\textbf {\bibinfo {volume} {81}},\ \bibinfo
  {pages} {144125} (\bibinfo {year} {2010})}\BibitemShut {NoStop}%
\bibitem [{\citenamefont {Kumar}\ and\ \citenamefont {Waghmare}(2010)}]{Kumar}%
  \BibitemOpen
  \bibfield  {author} {\bibinfo {author} {\bibfnamefont {A.}~\bibnamefont
  {Kumar}}\ and\ \bibinfo {author} {\bibfnamefont {U.~V.}\ \bibnamefont
  {Waghmare}},\ }\href@noop {} {\bibfield  {journal} {\bibinfo  {journal}
  {Phys. Rev. B}\ }\textbf {\bibinfo {volume} {82}},\ \bibinfo {pages} {054117}
  (\bibinfo {year} {2010})}\BibitemShut {NoStop}%
\end{thebibliography}%

\end{document}